# Exploring interlayer Dirac cone coupling in commensurately rotated few-layer graphene on SiC(000-1)

C. Mathieu[1], E. H. Conrad[2], F. Wang[2], J. E. Rault[1,*], V. Feyer[3,4], C. M. Schneider[3], O. Renault[5], N. Barrett[1]

[1] IRAMIS/SPEC/LENSIS, F-91191 Gif-sur-Yvette, France
[2] The Georgia Institute of Technology, Atlanta, Georgia 30332-0430, USA
[3] Peter Grünberg Institute (PGI-6), JARA-FIT, Research Center Jülich, 52425 Jülich, Germany
[4] NanoESCA beamline, Sincrotrone Trieste, Area Science Park, 34149 Basovizza, Trieste, Italy
[5] CEA, LETI, MINATEC Campus, F-38054 Grenoble Cedex 09, France
* Now at: Synchrotron-SOLEIL, BP 48, Saint-Aubin, F91192 Gif sur Yvette CEDEX, France

Corresponding author: claire.mathieu@cea.fr

Abstract:

We investigate electronic band-structure images in reciprocal space of few layer graphene epitaxially grown on SiC(000-1). In addition to the observation of commensurate rotation angles of the graphene layers, the k-space images recorded near the Fermi edge highlight structures originating from diffraction of the Dirac cones due to the relative rotation of adjacent layers. The 21.9° and 27° rotation angles between two sheets of graphene are responsible for a periodic pattern that can be described with a superlattice unit cells. The superlattice generates replicas of Dirac cones with smaller wave vectors, due to a Brillouin zone folding.

1. Introduction

The electronic properties of bilayer graphene have received high attention recently. Indeed, the rotation angle between the layers can lead to new physical properties. Bilayer graphene has been initially interpreted under the supposition of AB or Bernal stacking, which were the most common in graphitic materials. However, natural or synthetic crystals present different defects that can affect the stacking order (translation or/and rotation) in the c-direction. In twisted bilayer graphene, the two graphene layers are arbitrarily oriented, leading to possible interlayer interaction. It has been theoretically [1,2] and experimentally [3] shown that in such a system, interlayer interaction can occur at specific locations within the Brillouin zone (BZ). For example, this interaction can be responsible for Fermi velocity modifications [1], minigap opening at the BZ boundary [3] or Van Hove singularity observation [4]. Moreover, it has been demonstrated that the optical absorption spectrum can evolve by changing the rotation angle, regardless of the lattice commensurability [5-7]. This weak interaction, which modifies the electronic band structure, can potentially lead to new electronic properties of high interest.

Among graphene growth techniques, epitaxial graphene grown on C-terminated SiC substrates, noted SiC(000-1), offers the particular advantage of its rotational stacking that leads to very high mobilities [8,9]. As a result of the C termination, the graphene is decoupled from the substrate and subsequent layers show a disordered rotational stacking which effectively decouples adjacent graphene layers, each acting as almost free-standing graphene [10,11]. With the perspective of applications that imply more than one graphene layer for specific properties, it is therefore important to understand the electronic properties of such systems.

A lot of work has already been performed using Angle-Resolved PhotoEmission Spectroscopy (ARPES) to probe the band structure of epitaxial graphene on SiC(000-1) [12-14]. However, this type of epitaxial graphene develops domains at the micron scale that can present different thicknesses. To investigate the properties of a single domain, spatial resolution is therefore required. In ARPES, the spatial resolution is limited to a typical range of 50 to 100 μm, because of the photon beam size and the acceptance area of the spectrometer. It therefore often averages over several domains. An alternative way is the use of photoelectron emission microscopy (PEEM). This technique has made considerable progress in recent years thanks to instrumental innovations [15,16] and the availability of high brightness VUV and X-ray sources. Recent advances in Energy Filtered PEEM (EF-PEEM) allows the correlation of real space chemical and work function mapping with reciprocal space imaging of the complete band structure of micron-sized areas [17,18]. Compared to ARPES, PEEM measurements in the k-space directly produce an $I(k_x, k_y)$ image at a given kinetic energy, instead of $I(E,\theta)$ for a given azimuthal angle. For example, the Fermi surface can be acquired in a single shot experiment. This imaging mode allows the investigation of the full BZ, and therefore the visualization of new structures and replicas that can be easily missed in ARPES.
Here, we present an EF-PEEM study of micron scale regions of few layers graphene (FLG) epitaxially grown on C-terminated SiC substrates. The investigation of the band structure of FLG demonstrates the presence of different replicas due to BZ folding.

2. Experimental Methods

We present here experiments performed on epitaxial graphene samples grown on the C-terminated of the SiC substrates, noted hereafter SiC(000-1). The substrates were 6H conducting

SiC(000-1) from Cree, Inc. Before graphene growth, the substrates were first $H_2$ etched for 30 min at 1400°C. The samples were then grown in an enclosed graphite RF furnace using the confinement-controlled sublimation process (CCS) [19]. The growth was conducted at 1475°C for 20 min. Before PEEM measurements the samples were annealed at 500°C in vacuum to remove surface contamination. The surface cleanliness was verified using X-ray photoemission spectroscopy.

The experiments were performed on the TEMPO beamline [20] (SOLEIL, Saint-Aubin, FRANCE) and the NanoESCA beamline [21] (Elettra, Basovizza, ITALY). Both end stations were equipped with a NanoESCA EF-PEEM (Oxford Instruments/Omicron Nanoscience and FOCUS GmbH) [22,23]. The design combines a fully electrostatic PEEM column and an aberration compensated double-hemispherical analyzer as energy filter. Using a suitable transfer lens configuration, the first intermediate image in the back-focal plane can be accessed. Consequently, all the angular information of the photoelectrons emitted from the field of view is conserved for a given kinetic energy. In this configuration, the image no longer contains spatial information, but angular information from the emitted photoelectrons. Moreover, an adjustable field aperture placed in the first intermediate image plane can select a specific region in real space. By imaging the angular emission of the valence band electrons, the full band structure can therefore be mapped for a specific region of interest, down to few micrometers. For both experiments, the energy resolution was 200 meV. All measurements were carried out at room temperature.

3. Results

Using the EF-PEEM experiment at the TEMPO Beamline, the number of graphene layers was estimated from the core level intensities. By fitting the C *1s* core-level spectrum of different regions of interest (ROIs), we are able to determine the intensity ratio of the photoemission signal coming from graphene sheets and from the SiC substrate I(graphene)/I(SiC). Using the inelastic mean free path as a fitting parameter, this ratio gives the number of graphene layers in each ROI [17]. Two regions of interest of one and two monolayers were selected by closing the field aperture in the intermediate image plane down to ~7μm. Using the k-space imaging mode, the band structure of the two areas has been mapped. The constant energy cuts 1.3 eV below the Fermi level of single and bilayer graphene areas are displayed in figure 1a and 1b, respectively.

Figure 1a shows 6 Dirac cones, with the six-fold symmetry expected for a graphene monolayer. The constant energy cuts through the Dirac cones clearly show the extinction of intensity due to a matrix element effects, a well-known effect in graphene sheets [24,25]. Some fainter structures, rotated by ~20° relative to the main Dirac cones, are observed. They are probably due to electrons entering the instrument from areas outside the field aperture, from a domain with a different thickness and sheet orientation.

Figure 1b displays the band structure of a graphene bilayer. There are two sets of Dirac cones A and B. These are defined by their reciprocal lattice vectors (**a₁, a₂**) and (**b₁, b₂**) for layers A and B, respectively. The set of cones A is very intense, while the second set (B) is less bright and rotated by 21.9°, relatively to the set A. This angle has already been observed for graphene layers grown by CVD on polycrystalline Ni films [26] or epitaxially grown on SiC(000-1) substrates [27]. An additional set of cones appears inside the cone radius of the first Brillouin zone (FBZ) of sets A and B. For this third set of cones, the intensity extinction faces the Γ point, opposite to that for sets of cones A and B. These replicas can be attributed to diffraction effects between layer A and B, as explained below.

The stacking geometry of graphene layer A and B is described by the rotation angle of 21.9°. A sketch of the real-space lattice of both layers A (dashed blue) and B (red) is represented in the figure 2b. At this rotation angle, a periodic superstructure appears, giving rise to a new supercell. The vectors of this superlattice can be defined as an integer sum of the primary graphene vectors ($a_1$, $a_2$) and ($b_1$, $b_2$) for the adjacent layers A and B respectively. These vectors are therefore given by [28]:

$$L_1 = ma_1 + nb_1 = na_2 + mb_2 \quad (1)$$

$$L_2 = R\frac{\pi}{3} L_1 \quad (2)$$

for which n and m are integers. The superlattice is therefore define by the integer pair [m,n], which is, in this case [2,1], and the rotation angle θ is related to them by:

$$\cos\theta = \frac{1}{2} \times \frac{m^2 + n^2 + 4mn}{m^2 + n^2 + mn} \quad (3)$$

For specific angles, such as 21.9°, these commensurate graphene layers lead to band folding that should give rise to replicas inside the FBZ. This effect has been theoretically predicted for different commensurate rotations, such as 5.09° and 21.8° [5]. Figure 2b focuses on the energy cut of the bilayer, presented in figure 1b, on which the FBZ of layers A (blue) and B (red) are superimposed. The smaller hexagonal lattice (represented by dashed lines) is due to the band folding effect generated by the superlattice [2,1]. The dots indicate the position of the BZ at which replicas should appear inside the FBZ of layers A and B. The replicas observed on the figure 1b are therefore related to the superlattice structures and can be interpreted as Bragg diffraction of the electrons in the Dirac cones of one graphene layer by the periodicity of the adjacent layer. This conclusion is reinforced by the absence of such replicas in the case of a graphene monolayer. However, the six replicas that should appear around the Γ point, represented on figure 2b by unfilled dots, are not experimentally observed, possibly because of their weaker intensity.

In order to highlight inner replicas, generated by commensurate rotation, a second experiment has been performed at the NanoESCA beamline that possesses a micro-focused spot and thus provides a much higher photon flux within the field of view. This study was performed on another graphene multilayer sample epitaxially grown on a SiC(000-1) substrate. The constant energy cut, recorded at 1.5 eV below the Dirac point through a 5 μm field aperture, is shown in figure 3a. The reciprocal space horizon in the figure is 3.2 Å$^{-1}$ and the Γ point is at the center of the image. The image shows numerous sets of six Dirac cones, demonstrating the presence of several differently rotated graphene layers in the region of interest defined by the field aperture. Some rotations are of the order of a few degrees (9° for the green dashed arrow for instance), while others present higher rotation angles of 22° (blue arrow) and 27° (cone B), relative to the cone A. Moreover, some inner bands are visible inside Dirac cones B, due to graphene layers that present the same angular rotation in the stack. Figure 3b shows a cut perpendicular to KM direction at the cone B position. Three bands can be observed, and could be attributed to Bernal stacked graphene bilayer with a third misoriented layer called ABA', as it has been theoretically been predicted [29]. The band structure of this ABA' stacking presents a non-linear dispersion around the Dirac point. However, in our case, the energy resolution prevents to perform a more precise quantitative analysis of the π-bands.

Additional sets of cones, less intense, appear again inside the cone radius of the A and B FBZs. These replicas present an extinction of intensity facing the Γ point, as in the first experiment. Around the Γ point, some fainter replicas can be observed within the dotted circle. To improve the contrast of these structures, the contrast aperture in the diffraction plane has been closed around the center of the FBZ, leading to a k-space field of view of 1.2 Å$^{-1}$. In these conditions, a constant energy cut at 1.5 eV below the Fermi level has been acquired (figure 3c). Using longer acquisition times (60 instead of 20 seconds per image) the innermost features in reciprocal space can be clearly observed.

As the brighter set of Dirac cones (A and B) has a rotation angle of 27°, we will focus in the following only on their contribution to the main replicas. Figure 4a presents the real space atomic structure of the bilayer graphene with a 27° rotation angle. The superlattice, represented by the large diamond (green), is therefore defined by the integer pair [m,n]=[3,1] of equation 1. The related angle θ, calculated on the basis of equation (3) leads to an angle of 33°, which is equivalent by symmetry to 27° (60°-33°). Figure 4b represents the FBZs of layers A (dotted blue) and B (dashed red). They are superimposed on the fainter experimental replicas around the Γ point, displayed in figure 3b. If we extrapolate the hexagonal lattice as defined by this six-fold symmetry outwards and extend it to reach the primary reciprocal cells of the graphene layers it matches perfectly the two primary graphene lattices A and B. Thus, the small hexagonal lattice represents the folded BZ of the graphene bilayer A and B. As we pointed out, some multiple structures are also visible at the Dirac cone B position, due to ABA' stacking. It is therefore expected to observe inner structures in the replicas, which are clearly visible in figure 3c.

The dots superimposed on this folded BZ (figure 4b) indicate the position at which replicas should appear inside the FBZ of layers A and B for this specific angle. The sketch shows all the replicas that should be experimentally observed in the band structure. Moreover, looking closely to the K and K' point of the FBZs A and B (figure 3a), we can see other weaker inner Dirac cones, that can therefore be attributed to this Bragg diffraction effect of adjacent layers.

However, once again some replicas due to the [3,1] superlattice, that should have been experimentally observed, are not clearly distinguishable in the k-space image. They are represented in figure 4b by unfilled dots. The low intensity of these replicas may be due to their particular momentum value, but this aspect must be studied further.

4. Conclusions

Using k-space PEEM of the valence band of few-layer graphene on SiC(000-1) we have explored the full band structure over more than the FBZ for monolayer, bilayer and multilayer graphene. Simultaneous observation of several sets of Dirac cones shows the presence of different rotation angles of the stacked graphene sheets. We focused on graphene bilayer rotated by 21.9 and 27°. We have experimentally demonstrated the possible interaction between two adjacent sheets from the observation of Bragg diffracted Dirac cones, arising from the commensurate superlattice. This leads to replicas at different locations of the FBZs. Some faint replicas, observed around the Γ point, which to our knowledge have not yet been experimentally observed in photoemission, confirm the possibility to explore different replicas related to the BZ folding.

This study shows that EF-PEEM is an ideal instrument to explore the band structure of materials presenting regions of interest down to the micron-scale. The k-space spectroscopic imaging mode allows the observation of structures that might be missed in conventional ARPES experiments. However, until now, the EF-PEEM was still limited by an energy resolution of 200 meV at room temperature. This limiting factor has been recently improved, with a new instrumental design of the NanoESCA, allowing experiments with a 35 meV energy resolution at low temperature (35 K). This particular instrument will allow a better understanding of adjacent layer interactions, including the direct observation for all $k_x$, $k_y$ values of van Hove singularities and the possible minigap opening at specific location of the BZ. The study of these potential physical effects, combined with the crystallographic information that can be gained with this instrument, will allow a better understanding in view of new electronic applications.

Acknowledgement

We acknowledge SOLEIL for provision of synchrotron radiation facilities, and we would like to thank F. Sirotti and M. G. Silly, the TEMPO beamline staff.

References

[1] J. M. B. Lopes dos Sanos, N. M. R. Peres, A. H. Castro Neto, *Phys. Rev. Lett.* **2007**; 99, 256802.
[2] E. J. Mele, *Phys. Rev. B,* **2011**; 84, 2354391.
[3] T. Ohta, J. T. Robinson, P. J. Feibelman, A. Bostwick, E. Rotenberg, T. E. Beechem, *Phys. Rev. Lett.* **2012**; 109, 186807.
[4] G. Li, A. Luican, J. M. B. Lopes dos Santos, A. H. Castro Nteo, A. Reina, J. Kong, E. Y. Andrei, *Nat. Phys.* **2010**; 6, 109.
[5] P. Moon, M. Koshino, *Phys. Rev. B*, **2013**; 87, 205404.
[6] J. T. Robinson, S. W. Schmucker, C. Bogdan Diaconescu, J. P. Long, J. C. Culbertson, T. Ohta, A. L. Friedman, and T. E. Beechem, *ACS Nano*, **2013**; 7, 637.
[7] J. Campos-Delgado, G. Algara-Siller, C. N. Santos, U. Kaiser, J.-P. Raskin, *Small*, **2013**;9, 3247.
[8] C. Berger, Z. Song, X. Li, X. Wu, N. Brown, C. Naud, D. Mayou, T. Li, J. Hass, A. N. Marchenkov, E. H. Conrad, P. N. First, W. A de Heer, *Science*, **2006**; 312, 1191.
[9] M. Orlita, C. Faugeras, P. Plochocka, P. Neugebauer, G. Martinez, D. K. Maude, A.-L. Barra, M. Sprinkle, C. Berger, W. A. de Heer, M. Potemski, *Phys. Rev. Lett.* **2008**; 101, 267601.
[10] M. Sprinkle, D. Siegel, Y. Hu, J. Hicks, A. Tejeda, A. Taleb-Ibrahimi, P. Le Fèvre, F. Bertrand, S. Vizzini, H. Enriquez, S. Chiang, P. Soukassian, C. Berger, W. A. de Heer, A. Lanzara, E. H. Conrad, *Phys. Rev. Lett.* **2009**; 103, 226803.
[11] J. Hass, F. Varchon, J. E. Millán-Otoya, M. Sprinkle, N. Sharma, W. A de Heer, C. Berger, P. N. First, L. Magaud, E. H. Conrad. *Phys. Rev. Lett.* **2008**; 100, 125504.
[12] L. I. Johansson, S. Watcharinyanon, A. A. Zakharov, T. Iakimov, R. Yakimova, C. Virojanadara, *Phys. Rev. B*, **2011**; 84,125405.
[13] M. Sprinkle, J. Hicks, A. Tejeda, A. Taleb-Ibrahimi, P. Le Fèvre, F. Bertran, H. Tinkey, M. C. Clark, P. Soukiassian, D. Martinotti, J. Hass, E. H. Conrad, *J. Phys. D: Appl. Phys.* **2010**; 43, 374006.
[14] A. Tejeda, A. Taleb-Ibrahimi, W. de Heer, C. Berger, E. H. Conrad, *New J. Phys.* **2012;** 14, 125007.
[15] M. Escher, k. Winkler, O. Renault, N. Barrett, *J. Electron Sprectrosc. Relat. Phenom.* **2010**; 178-179, 303.
[16] A. Locatelli, E. Bauer, *J. Phys.: Condens. Mat.* **2008**; 20, 093002.

[17] C. Mathieu, N. Barrett, J. Rault, Y. Y. Mi, B. Zhang, W. A. de Heer, C. Berger, E. Conrad, C. Berger, O. Renault, *Phys. Rev. B*, **2011**; 83, 235436.
[18] J. E. Rault, J. Dionot, C. Mathieu, V. Feyer, C. M. Schneider, G. Geneste, N. Barrett, *Phys. Rev. Lett.* **2013**; 111, 127602.
[19] W. A. De Heer, C. Berger, M. Ruan, M. Sprinkle, X. Li, Y. Hu, B. Zhang, J. Hankinson, E. H. Conrad, *Proc. Natl. Acad. Sci.* **2011**; 108, 16900.
[20] http://www.synchrotron-soleil.fr/Recherche/LignesLumiere/TEMPO
[21] http://www.elettra.trieste.it/it/lightsources/elettra/elettra-beamlines/nanoesca/nanoesca.html
[22] M. Escher, N. Weber, M. Merkel, C. Ziethen, P. Bernhard, G. Schonhense, S. Schmidt, F. Forster, F. Reinert, B. Krömker, D. Funnemann, *J. Phys. Condens. Matter*, **2005**; 17, S1329.
[23] B. Krömker, M. Escher, D. Funnemann, D. Hartung, H. Engelhard, J. Kirschner, *Rev. Sci. Instrum.* 2008; 79, 053702.
[24] E. L. Shirley, L. J. Terminello, A. Santoni, F. J. Himpsel, *Phys. Rev. B*, **1995**; 51, 13614.
[25] I. Gierz, J. Henk, H. Höchst, C. R. Ast, K. Kern, *Phys. Rev. B*, **2011**; 83, 121408(R).
[26] A. Luican, G. Li, A. Reina, J. Kong, R. R. Nair, K. S. Novoselov, E. Y. Andrei, *Phys. Rev. Lett.* **2011**; 106, 126802.
[27] T. G. Mendes-de-Sa, A. M. B. Goncalves, M. J. S. Matos, P. M. Coelho, R. Magalhaes-Paniago, R. G. Lacerda, *Nanotechnology*, **2012** ; 23, 475602.
[28] E. J. Mele, *Phys. Rev. B*, **2010**; 81, 161405(R).
[29] S. Latil, V. Meunier, L. Henrard, *Phys. Rev. B*, **2007**; 76, 201402(R).

Figures

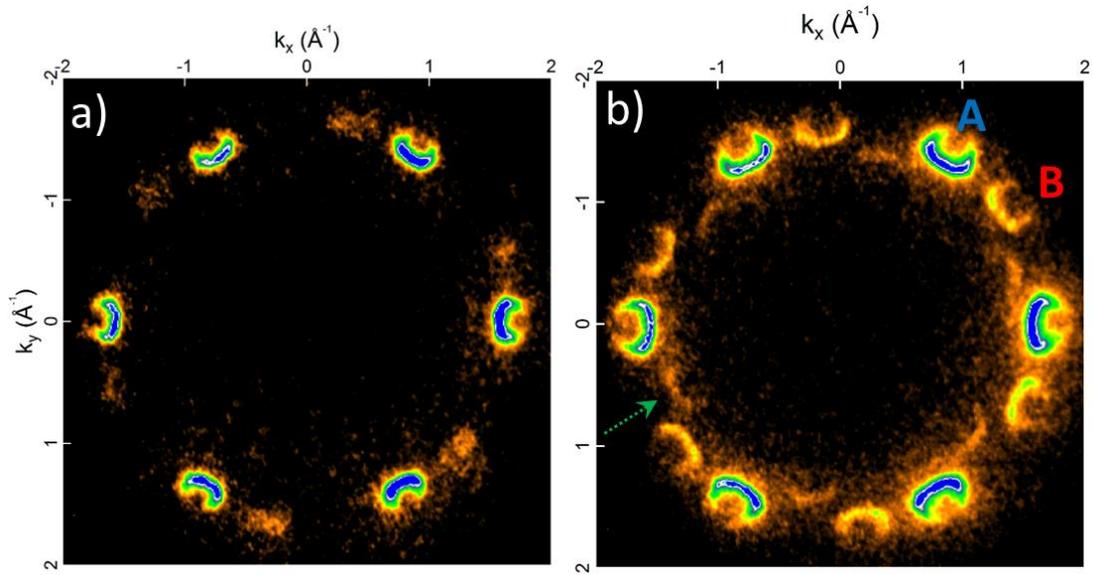

Figure 1. Constant energy cuts in the k-space, acquired at 1.3 eV below the Dirac point, with photon energy of 89 eV, for a monolayer (a) and a bilayer of graphene (b). On the image (b) two sets of cones are observed, noted A, B with a rotation angle relative to the A layer of 21.9°. One of the replicas with an inversed symmetry is shown by a dashed green arrow.

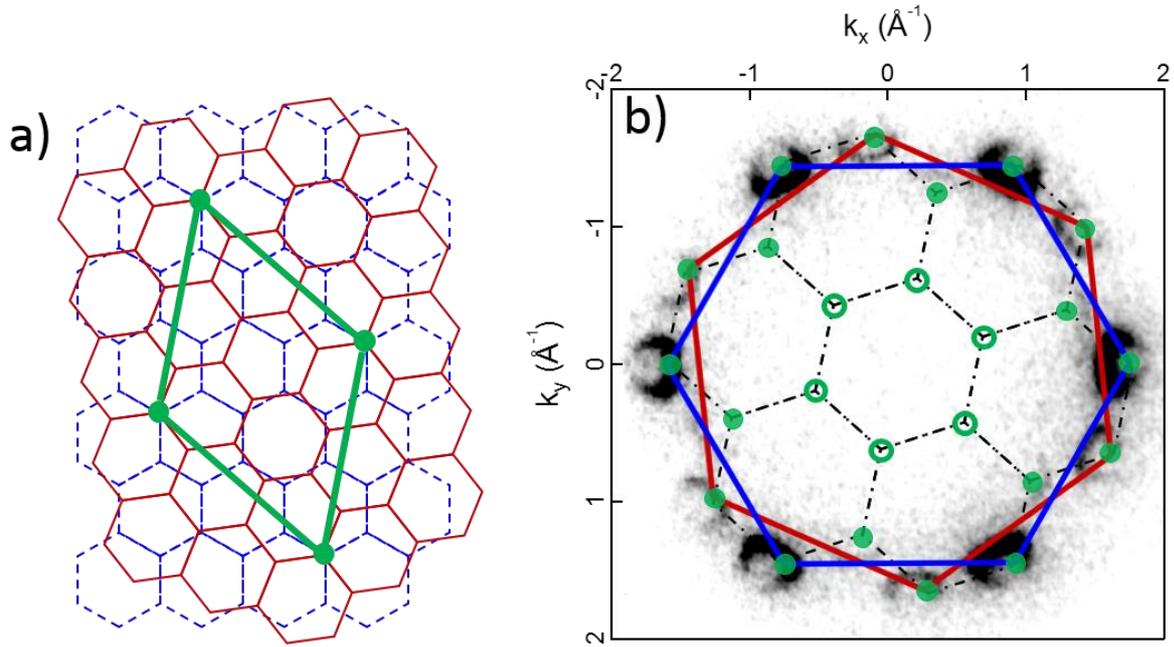

Figure 2. a) Atomic structure of the bilayer A and B with an angle rotation of 21.9°. The superlattice [2,1], formed at this specific angle, is superimposed on the image (green diamond). b) FBZs of layers A (blue) and B (red) with a rotation angle of $\theta=21.9°$, superimposed on the k-space image, displayed in figure 1b. The lattice composed by smaller hexagons is the folded BZ of this bilayer. The filled (unfilled) dots represent the location at which the replicas are experimentally observed (not observed) in the k-space image (fig. 1b).

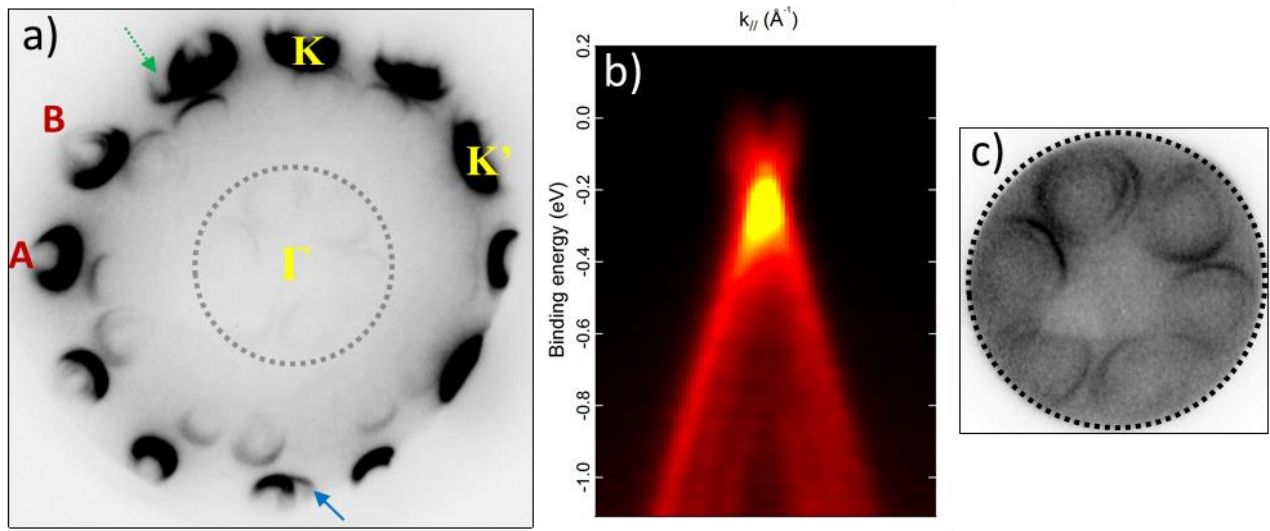

Figure 3. a) Constant energy cut in the k-space, acquired at 1.5 eV below the Dirac point, with photon energy of 51 eV. The field of view is 3.2 Å$^{-1}$. Two main intense sets of cones are observed, noted A, B with a rotation angle relative to the A layer of 27°. Other rotated sets of six Dirac cones are also observed, with angles of 9° (green dashed arrow) and 22° (blue arrow), relative to layer A. The Γ, K, K' points of the FBZ of layer A are also indicated. b) Band dispersion as a function of $k_{//}$ around the K point of Dirac cones B. c) The contrast of the fainter structures near the Γ point, as shown by the dotted circle in fig. 1a, has been improved by closing the contrast aperture in the diffraction plane, leading to a k-space field of view of 1.2 Å$^{-1}$ and by increasing the acquisition time by a factor 3.

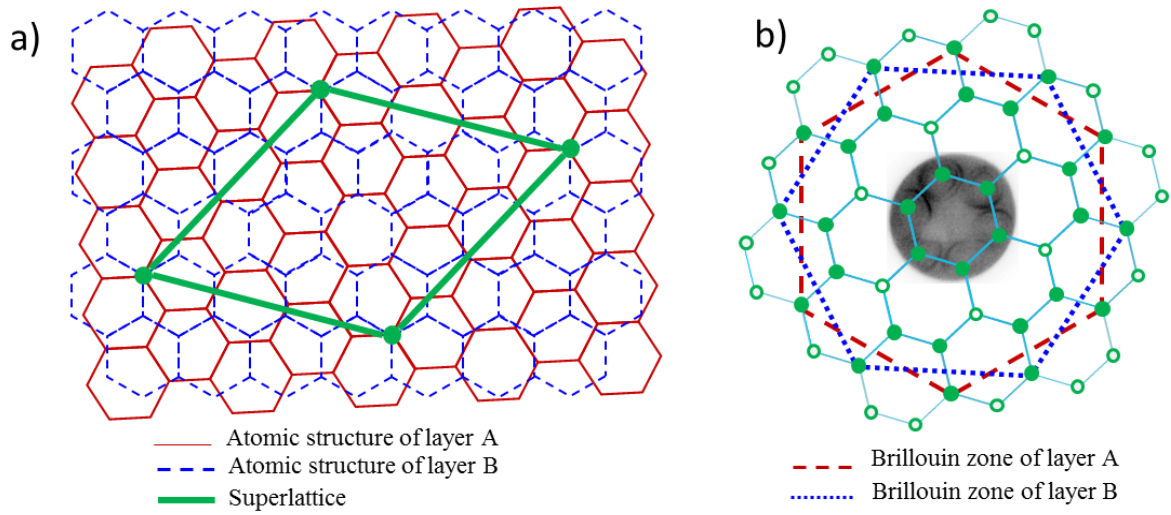

Figure 4. a) Atomic structure of the bilayer A and B with an angle rotation of 27°. The superlattice [3,1], formed at this specific angle, is superimposed on the image (green diamond). b) FBZs of layers A and B for a rotation angle of θ=27°. They are superimposed on the fainter structure showed in Fig. 1a. The lattice composed by smaller hexagons is the folded BZ of these two layers. The filled (unfilled) dots represent the location at which the replicas are experimentally observed (not observed) in the k-space image (fig. 3a).